\documentclass{article}
\usepackage{arxiv}

\usepackage[utf8]{inputenc} 
\usepackage[T1]{fontenc}    
\usepackage{hyperref}       
\usepackage{url}            
\usepackage{booktabs}       
\usepackage{amsfonts}       
\usepackage{nicefrac}       
\usepackage{microtype}      
\usepackage{graphicx}
\usepackage{natbib}
\usepackage{doi}

\usepackage{multirow}
\usepackage{amsmath,amssymb}
\usepackage{amsthm}
\usepackage{mathrsfs}
\usepackage[title]{appendix}
\usepackage{xcolor}
\usepackage{textcomp}
\usepackage{manyfoot}
\usepackage{booktabs}
\usepackage{algorithm}
\usepackage{algorithmicx}
\usepackage{algpseudocode}
\usepackage{listings}
\usepackage{lscape}
\usepackage{threeparttable}
\usepackage{tabularx}
\usepackage{longtable}
\usepackage{threeparttablex}

\newcolumntype{L}{>{\raggedright\arraybackslash}X}

\title{A Differential Index Measuring Rater's Capability in Educational Assessment}

\date{Februry 12, 2025}	

\author{You-Gan Wang \\
	The University of Queensland\\
		Brisbane, QLD, Australia, 4072\\
	\texttt{ygwanguq2012@gmail.com} \\	
	\And
	  Jinran Wu\\
	The University of Queensland\\
		Brisbane, QLD, Australia, 4072\\
	\texttt{jinran.wu@uq.edu.au} \\
    \And
    \href{https://0000-0002-5446-9758}{\includegraphics[scale=0.06]{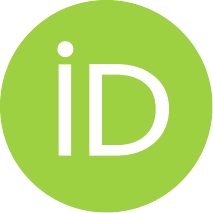}\hspace{1mm}Xuelan Qiu}\thanks{Corresponding author}\\
	Institute for Learning Sciences and Teacher Education\\
        Faculty of Education and Arts\\
	Australian Catholic University\\
	Brisbane, QLD, Australia, 4000\\
	\texttt{sherry.qiu@acu.edu.au} \\
}

\renewcommand{\shorttitle}{An Index for Rater's Capability}

\hypersetup{
pdftitle={Index Measuring Rater's Capability},
pdfsubject={psychometrics, educational.assessment},
pdfauthor={You-Gan Wang, Jinran Wu, Xuelan Qiu},
pdfkeywords={accuracy of raters, rating behaviors, item response theory models, facets model, hierarchical rater model, parameter estimation},
}

\begin{document}
\maketitle

\begin{abstract}
A rater's ability to assign accurate scores can significantly impact the outcomes of educational assessments. However, common indices for evaluating rater characteristics typically focus on either their severity or their discrimination ability (i.e., skills to differentiate between students). Additionally, these indices are often developed without considering the rater’s accuracy in scoring students at different ability levels. To address the limitations, this study proposes a single-value measure to assess a rater's capability of assigning accurate scores to students with varying ability levels. The measure is derived from the partial derivatives of each rater's passing rate concerning student ability. Mathematical derivations of the index under generalized multi-facet models and hierarchical rater models are provided. To ease the implementation of the index, this study develops parameter estimation using marginal likelihood and its Laplacian approximation which allows for efficient evaluation and processing of large datasets involving numerous students and raters. Simulation studies demonstrate the accuracy of parameter recovery using the approximate likelihood and show how the capability indices vary with different levels of rater severity. An empirical study further tests the practical applicability of the new measure, where raters evaluate essays on four topics:``family," ``school," ``sport," and ``work." Results show that raters are most capable when rating the topic of family and least capable when rating sport, with individual raters displaying different capabilities across the various topics.
\end{abstract}


In educational assessment, it is a common practice to evaluate competency based on human raters' judgments. However, the subjectivity of these judgments can introduce bias in the scores or ratings provided. To address this issue, researchers have proposed various item response theory (IRT) models that incorporate rater-specific parameters. These models can be broadly classified into two frameworks: the facets framework and the hierarchical rater model (HRM) framework.

In the facets framework \citep{linacre1989many}, raters act as independent judges, using their expertise to assign ratings ($0 =\mbox{incorrect or failure}$, $1 =\mbox{correct or pass}$) to students, essentially functioning as items in this framework. Let $Y_{nri}$ represent the ratings given by rater $r$ to student $n$ on item $i$. It is important to note that this study focuses on dichotomous ratings. Mathematically, the probability of $Y_{nri}=1$ in the facets model \citep{linacre1989many} can be expressed as:
\begin{equation} \label{facets_dicho}
P(Y_{nri}=1) = \frac{\exp(\theta_{n} - \delta_{i} - \eta_{r})}{1 + \exp(\theta_{n} - \delta_{i} - \eta_{r})},
\end{equation}
where $\theta_{n}$ represents the ability of student $n$, $\delta_{i}$ denotes the overall difficulty of item $i$, and $\eta_{r}$ indicates the severity of rater $r$. Equation (\ref{facets_dicho}) is referred to as the three-facets model (TFM) in this work since it encompasses three facets: student ability, item difficulty, and rater severity. 
In particular, the parameter $\eta_{r}$ characterizes the rater's overall passing rate.  It should be noted that no matter how severe or lenient the rating is (i.e. what $\eta_r$ value is), this parametric model assumes all judges have the correct order of passing each student, i.e., better students have a higher chance of passing. Therefore, this facet model assumes that all judges are competent in evaluating the students, even though the passing rates may vary.

In the HRM framework \citep{Patz2002hrm}, raters' scores are viewed as reflecting the underlying ``true'' category of an assessed task (e.g., an essay) through a two-level structure. The first level of the HRM specifies the process by which a rater determines the ``true'' category for the task. For instance, in the latent class HRM \citep{decarlo2005model,decarlo2011hrmlc}, a signal detection theory model is employed, which can be represented as follows:
\begin{equation} \label{hrm_l1}
P(Y_{nri}=1) = \textit{F} \left( c_{r} - a_{ri} \xi_{ni} \right).
\end{equation}
In this equation, $\xi_{ni}$ represents the latent ``true'' category of item $i$ for student $n$; $c_{r}$ is the criteria parameter for rater $r$; $a_{ri}$ is a slope parameter; and \textit{F} is a cumulative normal or logistic distribution. Equation (\ref{hrm_l1}) considers not only the rater's severity ($c_{r}$) but also their discrimination ability ($a_{ri}$), which measures the rater's capability to detect the latent categories of item $i$. The second level of the HRM involves the use of an item response theory (IRT) model, such as the Rasch model \citet*{rasch1960studies}, to model the probability of the ``true'' category of item $i$.

Although the original formulations of the TFM and HRM focus primarily on one or two rater characteristics—mainly severity—these frameworks provide a foundation for researchers to investigate raters' attributes more deeply. For example, rater consistency, which measures how consistently a rater assigns similar ratings to students with comparable ability levels \citep[e.g.,][]{uto2016rGRM,uto2020rGRM,qiu2022new}; rater centrality/extremity which measures the degree to which a rater excessively employs middle/extreme ratings \citep[e.g.,][]{jin2018new,qiu2022new,Wind2019, Wolfe2015,elliott2009}; and rater discrimination which reflects how effectively a rater distinguishes and segregates students across the ability continuum \citep{wu2017,Kim2021}.  


Our study focuses on the characteristic of a rater of assigning accurate ratings to students across different levels of ability, which is defined as ``capability" within this research. A capable rater is expected to assign higher scores to students with high ability and lower scores to those with lower abilities. Therefore, this characteristic implicitly involves two proficiencies of raters: first, the rater can identify the ability levels of students. In other words, the rater can accurately discriminate between students who have higher or lower abilities. Second, the rater can consistently assign appropriate scores to students, neither too severe nor too lenient. Lacking either proficiency might lead to inaccurate ratings. For example, a rater who is unable to discriminate student ability levels is likely to assign scores randomly and arbitrarily. An extreme case would be a rater who assigns one ``correct" score followed by one ``incorrect" score to the students. Similarly, a rater who consistently rates too severely or too leniently may still rate inaccurately because their ratings do not align with the true ability levels of the students.

Although many IRT models have been proposed to measure the severity or discrimination of a rater, using the parameters separately cannot capture the two aspects of a rater's characteristics because the parameters are assumed to be independent of each other. Moreover, as shown later in the paper, the proposed capability index is a complex combination of rater severity and discrimination parameters, derived based on students' ability levels. Hence, it is less optimal, if not inaccurate, to presume that the capability index can be obtained by simply calculating it using the severity and discrimination parameter estimates from the IRT modeling output.

The proposed rater capability index is a single value metric that can be used as a standardized measure to assess raters' ability to accurately assign ratings across varying levels of student ability. As shown in the methodology section, it is derived based on the partial derivatives of each rater's passing rate with respect to the student's ability and can be incorporated into both the facets and HRM frameworks. This single value index is useful for evaluating the performance of raters in the rating process, ensuring the reliability and validity of the ratings. This is especially important for high-stakes testing contexts, where the fairness and accuracy of ratings are critical concerns.

The main contribution of this work is threefold: 1) developing a single-value capability index, presenting the computation of the index under various IRT models, and providing mathematical proof for the computation of the index, 2) developing an estimation method utilizing the hierarchy likelihood (HL) and the Laplacian approximation which enables efficient evaluation of large datasets, and 3) investigating the performance of the index in identifying capable or incapable raters using simulation and empirical studies. The remainder of this paper is organized as follows. First, the definition of the capability index was introduced and the computation of the capability index was presented under different IRT models, including facets and HRM models. The mathematical computation of the index, underpinned by the proofs, was provided. Second, the parameter estimation methods for the index were introduced. Third, simulation studies were conducted to evaluate the parameter recovery of the proposed estimation method and the performance of the index in identifying raters with different capabilities under various conditions. The results were summarized. Fourth, an empirical example was provided to demonstrate the applications of the new index to measure raters' capability, and finally, the paper concluded with a summary and a comprehensive discussion. 

\section{Methodology}\label{methodology}
 In this section, we present the proposed capability index and show the mathematical computations under various IRT models. Unless otherwise stated, the ratings are binary and the notations are as defined in Section 1. 
 
 \subsection{Proposed Capability Index}
 Let $P(Y_{nri}=1|\boldsymbol{\Theta})$ be the probability of $Y_{nri}=1$ conditional on $\boldsymbol{\Theta}$ which is the collection of an unknown person, item, and rater parameters. It can be defined as
\begin{equation} \label{glm} 
P(Y_{nri}=1|\boldsymbol{\Theta})= F(S_{nri}),
\end{equation}
where $S_{nri}$ is a linear or nonlinear combination of person, item, and rater variables. For generality, define $S_{nri} = g(\theta_n, \delta_i, \eta_r)$ where function $g$ will allow synergistic interaction between the student ability and the rater capability,
as the case in the HRM. $F$ is a probability distribution satisfying $F(-\infty)=0$ and $F(\infty)=1$.  For binary classification, five link functions can be used as $F$, namely, $logit$, $probit$, $cauchit$, (corresponding to logistic, normal, and Cauchy CDFs respectively), $log$, and $cloglog$ (complementary log-log). Based on Equation (\ref{glm}), $P(Y_{nri}=0|\boldsymbol{\Theta})= 1-F(S_{nri})$. For dichotomous data, the mean value (denoted as $\mu_{nri}$) can be calculated as 
$\mu_{nri}= E(Y_{nri}=1|\boldsymbol{\Theta})= \sum_{k=0}^1 kP(Y_{nri}=1|\boldsymbol{\Theta}) = P(Y_{nri}=1|\boldsymbol{\Theta})$. 

In this study, we define the partial derivative of $\mu_{nri}$ with respect to the person parameter $\theta_n$, as a measure of the capability of rater $r$. When there is no confusion, the item subscript $i$ is dispensed with. Hence, the definition can be mathematically expressed as,
\begin{equation} \label{kappa} 
\kappa_r (\theta) = \frac{\partial \mu_{nr}/\partial \theta}{\Delta}
= \frac{\partial P(Y_{nr}=1|\boldsymbol{\Theta})/\partial \theta}{\Delta}
=   \frac{\partial S_{nr}/\partial \theta}{\Delta} f(S_{nr}),
\end{equation}
where $f(\cdot)$ is the density function of $F$, and the other terms are defined as above. $\Delta$ is a scale constant to standardize the capability index to ensure the ``perfectly capable" rater has the capability value 1. It is defined as the supremum over all the raters' parameter space as:
\begin{equation} \label{delta}
   \Delta = \sup_r \int_{-\infty}^{+\infty} \frac{\partial \mu_{nr}}{\partial \theta}  \phi(\theta) d \theta.
\end{equation}

In general,  $\kappa_r$ is a function of $\theta$. Under the assumption that $\theta_n$ follows a distribution $\phi(\theta)$ (e.g., standard normal distribution), a single value, representing the overall capability for the rater $r$, is defined as
\begin{equation} \label{kappa_bar} 
\bar{\kappa}_r = \int_{-\infty}^{+\infty} \kappa_r(\theta)  \phi(\theta) d \theta.
\end{equation}

The capability index $\kappa$ can serve as an indicator of a rater's proficiency for at least two significant reasons. Firstly, the numerator ${\partial \mu_{nr}/\partial \theta}$ or $\partial P(Y_{nr}=1|\boldsymbol{\Theta})/\partial \theta$ within Equation (\ref{kappa}) mathematically represents the degree of sensitivity in the probability of assigning a score of $1$ by rater $r$ concerning the changes in the rates' abilities. Thus, a higher value of $\kappa_r$ reflects the heightened sensitivity of rater $r$ in distinguishing between latent ability levels, ultimately leading to more accurate ratings (given severity parameters are the same).

Let's consider two students, Student 1 and Student 2, where Student 2's ability $\theta_2$ is slightly higher than Student 1's, indicated by a small positive difference $\zeta$ between them. In the context of a reasonable rater, one would anticipate that the probability $Y_{2r} = 1$ should be slightly greater than the probability $Y_{1r} = 1$, which is to say $\mu_{2r} \geq \mu_{1r}$. If we observe $\mu_{2r} = \mu_{1r}$, it suggests that rater $r$ is relatively unresponsive to changes in student ability. Conversely, when $\mu_{2r} \gg \mu_{1r}$, it signifies a high sensitivity to these changes. The extent of this change is quantified by the ratio $(\mu_{2r} - \mu_{1r}) / \zeta$, which ultimately converges to the partial derivative with respect to $\theta$.

Second, how capable rater $r$ can depend on the information the rater has regarding different levels of students. As shown in Equation (\ref{info_kappa}), in the binary case that is considered in this study, the Fisher information of the rater $r$ with respect to the student ability $\theta$ is proportional to the square of $\kappa_r$. Simply put, the larger amount of information, the more capable the rater to assigning the appropriate scores for the students, or vice versa. As such, $\kappa$ can be used to represent the capability of a rater,
\begin{equation} \label{info_kappa} 
 \mathcal{I}_{nr}(\theta) 
   = E \left[\frac{Y_{nr}}{\mu_{nr}} - \frac{1-Y_{nr}}{1- \mu_{nr}}\right]^2  \left(\frac{\partial \mu_{nr}}{\partial\theta} \right)^2
  = 
   \frac{1}{\mu_{nr} (1-\mu_{nr})} \left(\frac{\partial \mu_{nr}}{\partial\theta} \right)^2 \propto \kappa_r ^2.
\end{equation}

\subsection{Computing the Capability Index under Different Models}
We now show how the capability index  $\bar{\kappa}$ can be computed with different models and provide the underpinned mathematical proofs in the Appendices to save space.

\noindent {\bf Example 1. The three-facet model}

It can be shown that the TFM is a special case of Equation (\ref{glm}) using $logit$ link function in the $F$ function and defining $S_{nri} = \textstyle   \theta_n-\delta_i-\eta_r$. Here we assume that both 
mean $\theta_n$ and mean $\delta_i$ are 0 and the mean $\eta_r$ is denoted as $\alpha$ (which can also be regarded as the intercept).  Then, we have,
$$
\mu_{nri}= \frac{\exp\left(\textstyle   \theta_n-\delta_i-\eta_r\right)}{1+\exp\left(  \theta_n-\delta_i-\eta_r\right)}.
$$

Following the definition of capability index (i.e., Equation (\ref{kappa})), the numerator becomes
\begin{equation} \label{specialcase_facets_numerator}
\begin{split}
\frac{\partial \mu_{nr}}{\partial\theta_n} &= \frac{ \exp\left(\textstyle  \theta_n-\eta_r\right) }
{[1+\exp\left(  \theta_n-\eta_r\right)]^2} \\
& =    P(Y_{nri}=0|\boldsymbol{\Theta}) P(Y_{nr}=1|\boldsymbol{\Theta}) \\
& = \mu_{nr}(1-\mu_{nr}).
\end{split} 
\end{equation}
For the denominator of Equation (\ref{kappa}), by assuming $\theta$ follows standard normal distribution (denoted as $x$), we have
\begin{equation} \label{delta_tfm_denominator}
 \Delta = \sup_r \int_{-\infty}^{+\infty} 
 \frac{\exp(x)}{[1+\exp( x)]^2} \phi(x+\eta_r) dx =
 \int_{-\infty}^{+\infty} 
 \frac{\exp(x)}{[1+\exp( x)]^2} \phi(x) dx,
\end{equation}
where the supremum is achieved at $\eta_r  =0$ and the integration results in a constant value of $\Delta\approx 0.2066$. Readers interested in the proofs of the results can refer to Appendix A. 
So, the capability index under the TFM is 
\begin{equation} \label{facets_kappa}
\kappa_{r}^{TFM} = \frac{\partial \mu_{nr}/\partial \theta}{\Delta} = \frac{ \mu_{nr}(1-\mu_{nr}) }{ \Delta } = \approx 4.840 \mu_{nr}(1-\mu_{nr}), 
\end{equation}
and the overall capability for rater $r$ becomes
\begin{equation} \label{facets_kappa_bar}
\bar{\kappa}_r^{TFM} = \Delta^{-1} \int_{-\infty}^{+\infty} \mu_{nr}(1-\mu_{nr}) \phi(\theta) d \theta \approx 4.840 \int_{-\infty}^{+\infty} \mu_{nr}(1-\mu_{nr}) \phi(\theta) d \theta.
\end{equation}

\begin{figure}[htp]
    \centering
    \includegraphics[width=\textwidth]{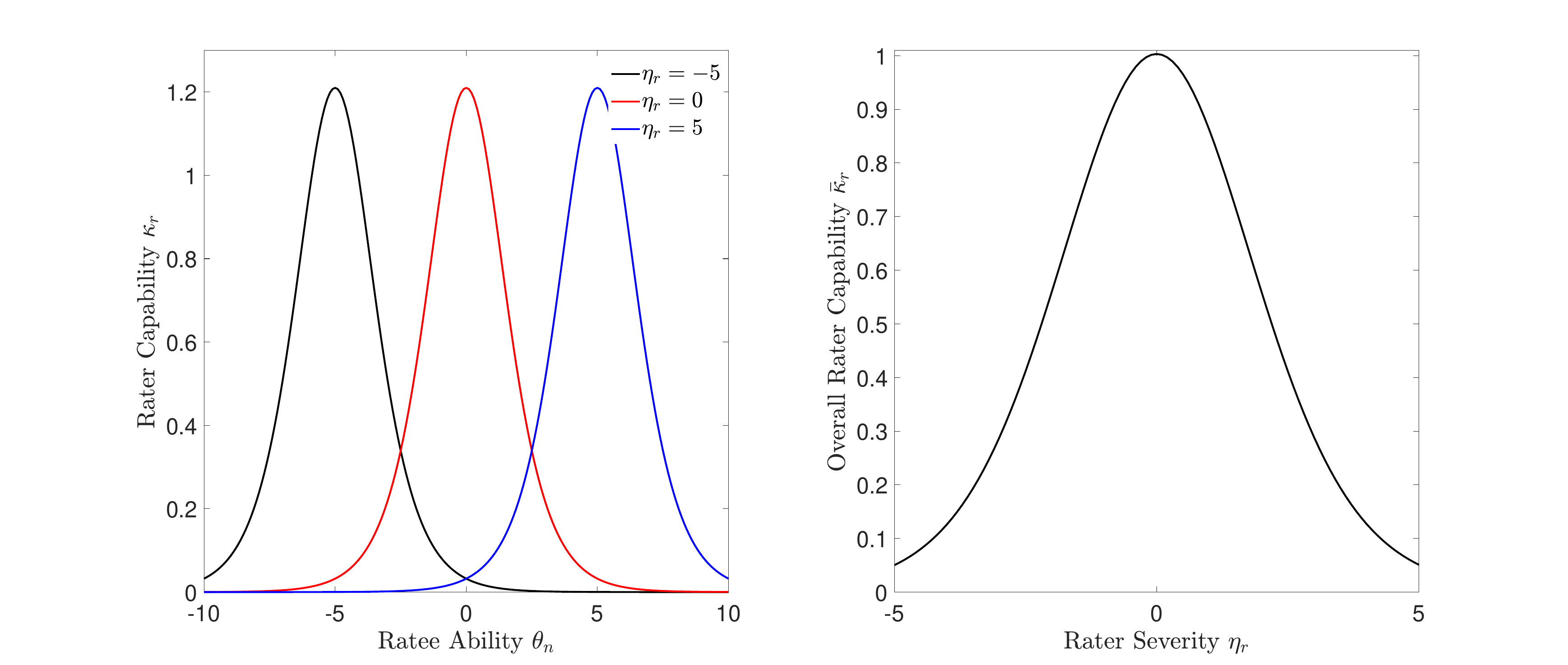}
    \caption{The capability index $\kappa$ (left) and overall capability index $\bar{\kappa}$ (right) under the three-facets model}
    \label{example1}
\end{figure}

For a deeper understanding of the capability index, Figure~\ref{example1} illustrates the behavior of $\kappa$ (left panel) and $\bar{\kappa}$ (right panel) using the TFM for three raters, each with distinct severity values and $\delta=0$. It's important to note that altering $\delta$ doesn't change the shapes of the curves but only shifts them along the x-axis. In the left panel, which presents the relationship between $\kappa$ and students' ability, we can observe a normal distribution of capability values for each rater, implying that raters with equal severity can possess varying levels of capability. Furthermore, the curves for the three raters reveal that those with differing severity levels can effectively differentiate between students of varying abilities, reaching their peak capability when $\theta_n = \eta_r$. For instance, for students with $\theta=5$, raters with $\eta=5$ will exhibit greater capability compared to other raters in assessing these high-ability students. Moreover, the $\kappa$ values for capability remain constant at points where $\theta_n = \eta_r$. The right panel depicts $\bar{\kappa}$ by integrating $\theta$ across $\mathcal{N}(0,1)$, consequently highlighting that raters having $\eta=0$ achieve the highest overall capability (i.e., 1). These outcomes provide valuable insights into rater recruitment strategies—when considering students from a normally distributed population, recruiting raters with $\eta$ values close to 0 generally offers higher information for the population on average. However, it is important to note that these raters may not yield the highest precision for all examinees. For instance, raters with $\eta = 0$ may exhibit lower discrimination for high-ability examinees, highlighting the need to consider the match between rater characteristics and specific examinee groups.

Based on the TFM formulation and the illustration in Table 1, it's apparent that identical passing rates – whether applied to high or low-ability students – will yield equivalent $\eta_r$ values, leading to the same $\kappa$ and $\bar{\kappa}$ values under the TFM. Nevertheless, this outcome isn't ideal in practical scenarios, as raters who pass more high-$\theta_n$ students should be recognized as more capable. The TFM, by implying that raters are equally skilled or capable if their $\eta_r$ values match, lacks the capacity to handle such diversity among raters. Consequently, the GMF model becomes more relevant in addressing these limitations.

\noindent {\bf Example 2. The generalized multi-facets model (GFM)}

The formulation of the GMF model with the rater discrimination parameter can be expressed as~\cite{uto2020rGRM}:
\begin{equation} \label{gmf}
P(Y_{nri}=1)= \frac{\exp\left(\textstyle \rho_r \theta_n - \delta_i -\eta_r \right)}{1+\exp\left( \rho_r \theta_n- \delta_i -\eta_r \right)},
\end{equation}
 in which, for parameter identifiability,  both the mean $\theta_n$ and the mean $\delta_i$ are assumed to be 0 and the mean $\eta_r$ is denoted as $\alpha$ (which can also be regarded as the intercept) as in Example 1, 
 and others are defined as in Equation (\ref{facets_dicho}). The GMF can be shown to be the special case of Equation (\ref{glm}) by using the $logit$ link function in the $F$ function and defining $S_{nri}=\rho_r\theta_n-\eta_r-\delta_i$. In this work, we reparameterize $\theta_n$ by adding a scaling parameter $\sigma$ to $\theta$, making $S_{nri}$ mathematically equivalent to be $\rho_r\sigma\theta'_n-\eta_r-\delta_i $ with $\mbox{var}(\theta'_n)=1$. This standardization procedure allows students' ability estimates under the GMF and TFM to be comparable.   

The boundary of $\rho_r$ in the GMF model is another noteworthy aspect. Unlike the item discrimination parameter found in certain IRT models (e.g., the 2PL model) that lacks an upper limit, the maximum value for $\rho_r$ in the GMF is confined to 1. This limitation arises from both theoretical and practical considerations. Specifically, referencing Equation (\ref{gmf}), the corresponding Fisher information is $E \partial^2 L_{nri} (\theta)/\partial \theta^2= \rho_r^2 \sigma^2\mu_{nri}(1-\mu_{nri})$, signifying that the information is proportionate to $\rho_r^2$. Let's consider two raters with a $\rho_2/\rho_1$ ratio of 10. Unrestricted values would imply that the information contributed by rater 2 is 100 times that of rater 1, introducing instability in estimation and compromising interpretability in practical applications. Conversely, bounded values ensure that the maximum $\rho_r$ is 1, offering more stable estimations and enhancing interpretability.

For the GMF model, we have 
\begin{equation} \label{generalized_facets_kappa}
\kappa_{r}^{GMF} = \frac{ \rho_r \sigma P(Y_{nr}=0|\boldsymbol{\Theta}) P(Y_{nr}=1|\boldsymbol{\Theta}) } { \Delta } = \Delta^{-1}\sigma \rho_r\mu_{nr}(1-\mu_{nr}),
\end{equation}
where
$$
  \mu_{nr}= \frac{\exp\left(\textstyle \rho_r\sigma\theta'_n-\eta_r \right)}{1+\exp\left( \rho_r\sigma\theta'_n-\eta_r \right)}.
$$
and  
$$
\begin{array}{cc}
\begin{aligned}
 \Delta & =  \int_{-\infty}^{+\infty} 
 \frac{\exp(x)}{[1+\exp( x)]^2} \phi(x/\sigma) dx \\
& \approx \frac{1}{4}\sqrt{\frac{2\sigma^2}{2+\sigma^2}}.
\end{aligned}
 \end{array}
$$
Readers interested in the proofs can refer to Appendix B.

Hence, the overall capability for rater $r$ becomes
\begin{equation} \label{generalized_facets_kappa_bar}
\begin{array}{cc} 
\begin{aligned}
\bar{\kappa}_r^{GMF}& =  \Delta^{-1}\sigma \rho_r\int_{-\infty}^{+\infty} \mu_{nr}(1-\mu_{nr}) \phi(\theta) d \theta.\\
&\approx 4 \rho_r \sqrt{\frac{2+\sigma^2}{2}} \int_{-\infty}^{+\infty} \mu_{nr}(1-\mu_{nr}) \phi(\theta) d \theta.
\end{aligned}
\end{array}
\end{equation}

\begin{figure}[htp]
    \centering
    \includegraphics[width=\textwidth]{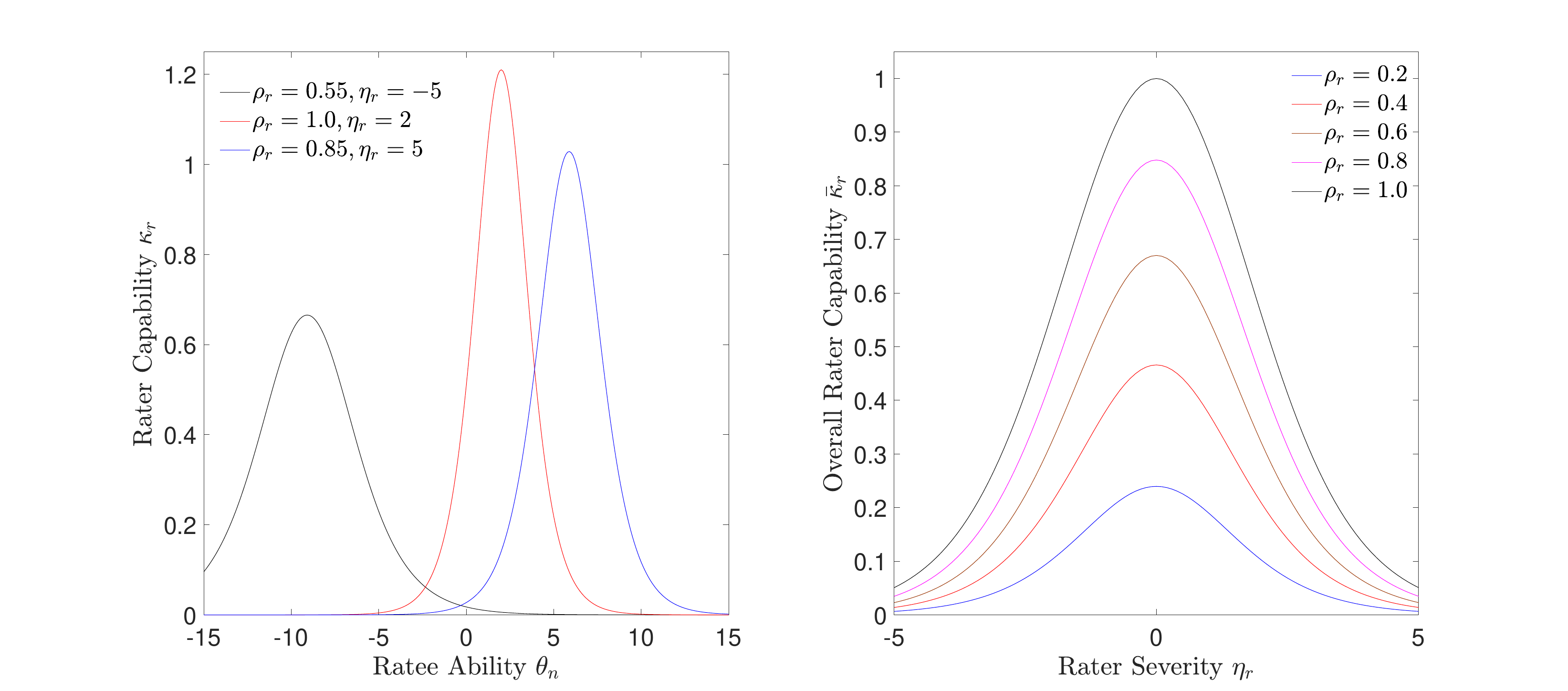}
    \caption{The capability index $\kappa$ (left) and overall capability index $\bar{\kappa}$ (right) under generalized multi-facets model}
    \label{example2}
\end{figure}

As shown in Equations (\ref{generalized_facets_kappa}) and ~(\ref{generalized_facets_kappa_bar}), $\kappa_r$ and $\bar{\kappa}_r$ under the GMF are the functions of rater discrimination parameter $\rho_r$, as well as rater severity parameter $\eta_r$. To visualize the effects of $\rho_r$ on $\kappa_r$, the left panel in Figure~\ref{example2} plots three raters with different combinations of $\rho$ and $\eta$ parameters. Compared to Figure~\ref{example1}, the spread of the normal distributions and the peaks vary across the three curves. The larger the $\rho$, the steeper the curve, and the higher the $\kappa_r$ value. The right panel in Figure~\ref{example2} shows that under the GMF the raters are unnecessarily equally capable even if their severity is the same. 

\noindent {\bf  Example 3. Probit Models}

 If we assume the observation by rater $r$ is perturbed by a random noise that is normally distribution $N(0, \sigma_r^2)$,
 the underlying score $\theta_n$ is basically perceived as $Z_{nri}= \theta_n +\sigma_r\epsilon_r$, where $\epsilon_r$ follows a standard normal with mean 0 and variance $1$  \citep{baker2004irt}. Here $\sigma_r$ is the scale of the perturbation.  The larger $\sigma_r$, the more damage the noise will do to the score. 
  Therefore,  $\sigma_r$ is also a measure of accuracy or capability for rater $r$, but in a reverse way. The observed binary outcome from rater $r$ 
  can be viewed as an indicator for whether $Z_{nri}$ is above the threshold $\alpha_r$,
\begin{equation} \label{probit}
  Y_{nri} = I(Z_{nri} -\alpha_r >0)=I(\theta_n +\sigma_r\epsilon_r-\alpha_r >0).
\end{equation}

The probit model can be shown to be the special case of Equation (\ref{glm}) by using the $probit$ link function in the $F$ function and defining $S_{nri} =  (\theta_n  - \alpha_r)/\sigma_r$ with $\eta_r = \alpha_r/\sigma_r$ which is the severity parameter.

Under the probit model, we have 
$$
\mu_{nri}= \Phi(S_{nri}),
$$
and
$$
\frac{\partial \mu_{nri}}{\partial\theta}   =  \phi(S_{nri})/\sigma_r, 
$$
where $\Phi$ and $\phi$ are the cumulative and density distribution functions of a standard normal random variable, respectively.
 
The $\kappa_r$ and $\bar{\kappa}_r$ under the probit model can be computed as:
\begin{eqnarray*}
\kappa_r^{PROBIT} & = &  \frac{1}{\sqrt{2\pi}\Delta\sigma_r} \exp(- S_{nri}^2/2), \\ 
\bar{\kappa}_r^{PROBIT} & = &    \frac{1}{\sqrt{2\pi}\Delta\sigma_r}\int_{-\infty}^{+\infty} \exp(- S_{nri}^2/2) \phi(\theta) d \theta
  = \frac{\rho_r}{\sqrt{2\pi}\Delta} \exp\left\{-\frac{(1-\rho_r^2)\eta_r^2}{2}\right\},
\end{eqnarray*}
in which  $\rho_r = 1/\sqrt{1+\sigma_r^2}$. As aforementioned, $\sigma$ represents the capability, 
 and $\rho$ is a simple inverse function of $\sigma$. Interestingly, $0\leq \rho_r\leq 1$. Therefore, the $\rho_r$ under the probit model has a similar meaning to the one in the GMF model. 

As shown in Appendix C, $\Delta= \phi(0)=1/\sqrt{2\pi}$ under the probit model. Hence, the capability and the overall capability index for this parametric model are 
\begin{eqnarray*}
\kappa_r^{PROBIT} & = &  \frac{1}{\sigma_r}\exp(- S_{nri}^2/2), \\ 
\bar{\kappa}_r^{PROBIT} & = &   \rho_r \exp\left\{-\frac{(1-\rho_r^2)\eta_r^2}{2}\right\},
\end{eqnarray*}
respectively, for rater $r$.

\begin{figure}[htp]
    \centering
    \includegraphics[width=\textwidth]{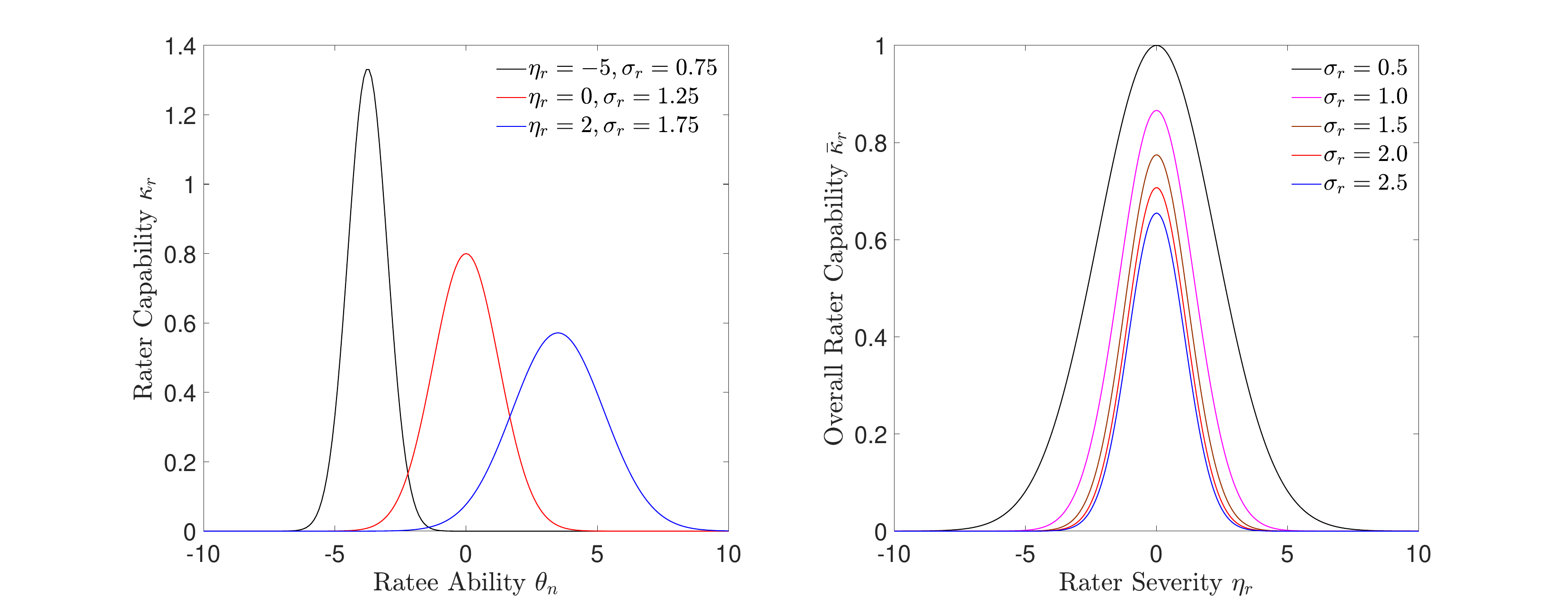}
    \caption{The capability index $\kappa$ (left) and overall capability index $\bar{\kappa}$ (right) under the probit model}
    \label{example3}
\end{figure}

The left panel of Figure~\ref{example3} plots $\kappa_r$ using $\sigma_r=0.75, 1.25, 1.75$ (corresponding $\rho_r=0.80, 0.62, 0.50$) with $\eta=-5, 0, 2$. It shows that the smaller the $\sigma_r$, the smaller the random noise, and the higher the $\kappa_r$ value. Like the GMF, raters with the same $\eta$ values are unnecessarily equally capable under the probit model, as the right panel of Figure~\ref{example3} shows.

It shall be noted that $\bar{\kappa}_r^{PROBIT}$ is an increasing function in $\rho_r \in [0, 1]$, and hence a decreasing function of $\sigma_r$ because of 
$$
\frac{\partial \bar{\kappa}_r^{PROBIT}}{\partial \rho_r}=(1+\rho_r^2 \eta_r^2) \exp\left\{-\frac{(1-\rho_r^2)\eta_r^2}{2}\right\}>0.
$$

Also, clearly, $\bar{\kappa}_r^{PROBIT}$ is a decreasing function of $|\eta_r|$.  This indicates that the capability increases when $\rho_r$ is larger and rater capability decreases when $\eta_r$ further deviates from 0.

\noindent {\bf  Example 4. Hierarchical Rater Models}

As mentioned earlier, the HRM \citep[e.g.,][]{decarlo2011hrmlc, qiu2022new} consists of two levels where Level 1 is a rater model and Level 2 is an IRT model. For binary data, there are two latent distributions for the true ratings: $\xi_{ni}=0$ and $\xi_{ni}=1$. 

The HRM can be shown to be the special case of Equation (\ref{glm}) by defining a mixture distribution with  
$$
  \mu_{nri}= F(\theta_n,\eta) = \sum_{k=0}^{1} F_k  F_{2,k},
$$ 
where $F_0 =P(\xi_{ni}=0)$ and $F_1=P(\xi_{ni}=1)$ are the probability function of the true rating $\xi_{ni}=0$ and $\xi_{ni}=1$, respectively, and $F_{2,k}$  is the probability of $Y_{nri}=1$ conditional on $\xi_{ni}=k$ ($k=0,1$). 
The implied marginal probability $P(Y_{nri}=1)$ is  $F(\theta_n,\eta)= E_\xi [E_Y(Y_{nri}=1|\xi_{ni})]= E_\xi F_{2,\xi_{ni}} = F_1 F_{2,0} + (1-F_1) F_{2,1}$. This indicates that  
 $$
\frac{\partial \mu_{nri}}{\partial\theta}   = (F_{2,1} -F_{2,0}) \frac{\partial F_1}
{\partial\theta}.
$$
The overall capability for rater $r$ becomes 
$$
\bar{\kappa}_r^{HRM} =  \Delta^{-1} (F_{2,1} -F_{2,0}) \int_{-\infty}^{+\infty} \frac{\partial F_1} {\partial\theta} \phi(\theta) d \theta.
$$
As shown in Appendix D,
$$
\Delta = \int_{-\infty}^{+\infty} \frac{\partial F_1} {\partial\theta} \phi(\theta) d \theta.
$$
Therefore, we have
\begin{equation} \label{specialcase_hrm_res}
\bar{\kappa}_r^{HRM} = \kappa_r^{HRM} = F_{2,1} - F_{2,0},
\end{equation}
where $F_{2,1} = P(Y_{nrj}=1|\xi_{ni}=1)$ and $F_{2,0} = P(Y_{nrj}=1|\xi_{ni}=0)$. 

\begin{figure}
    \centering
    \includegraphics[width=0.5\textwidth]{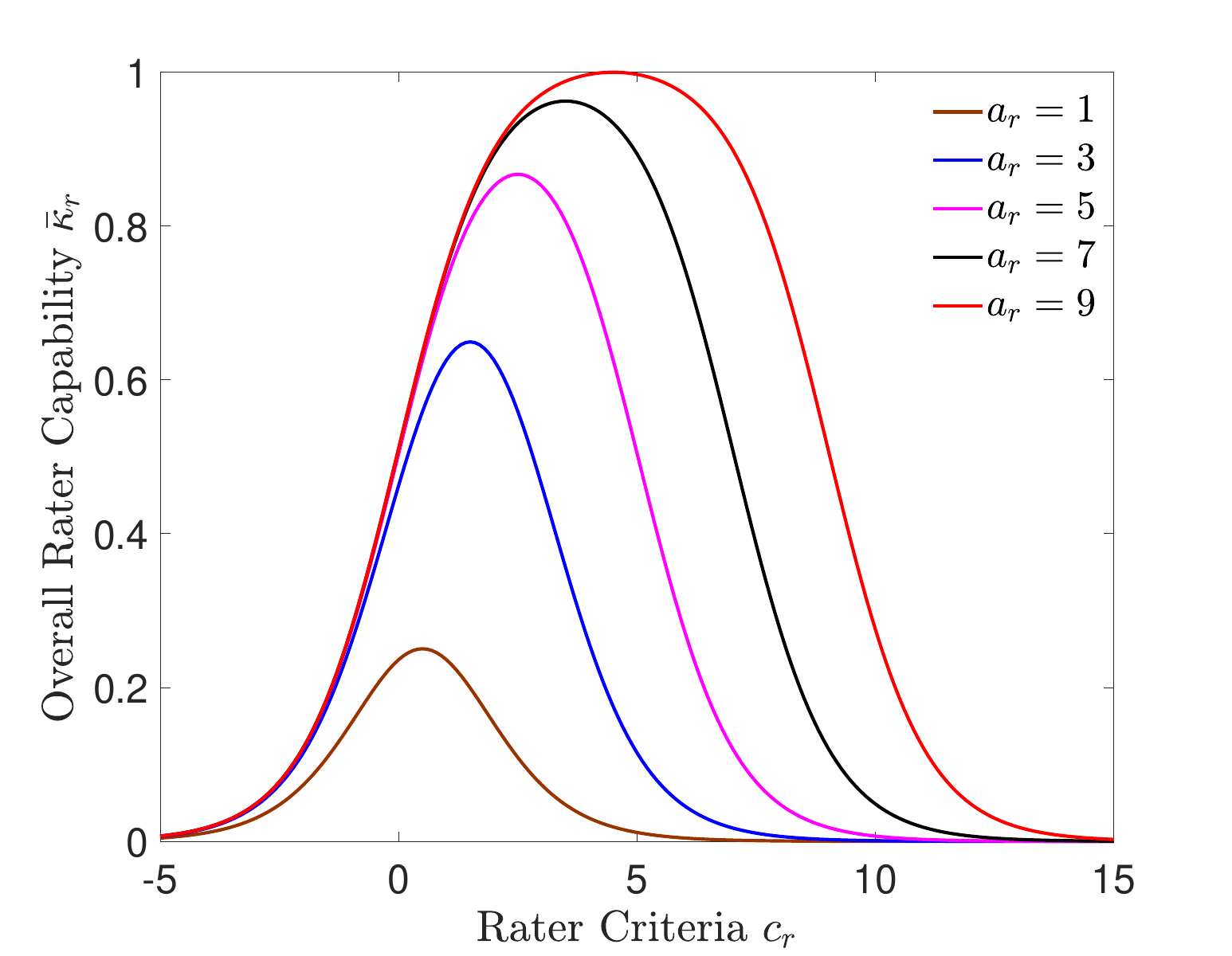}
    \caption{The overall capability index $\bar{\kappa}$ under the hierarchical rater models}
    \label{example4}
\end{figure}

Figure~\ref{example4} shows the capability index under the HRM with different slope parameters $a_r=1, 3, ..., 9$. Since $ \kappa_r^{HRM} = \bar{\kappa}_r^{HRM}$, only the $\bar{\kappa}_r^{HRM}$ is plotted. It can be seen that the larger the $a$ parameter, the higher the $\bar{\kappa}_r^{HRM}$ value.

\section{Parameter Estimation}\label{parameter-estimation}

This section introduces the estimation method we specifically propose to estimate the capability index. We first introduce the most widely used estimation method in IRT, discuss its limitations for the implementation of the capability index, and introduce the new estimation method. Due to time constraints, this work focuses on implementing the index under the GMF model because of its generability and popularity. 

Following Equation (\ref{glm}), the log-likelihood (LL) function for the parameters can be written as
\begin{equation}
\begin{split}
LL = &\sum_{n=1}^{N}\sum_{r=1}^{R}\sum_{i=1}^{I} \log \{P(Y_{nri}=y_{nri})\}\\
=&\sum_{n=1}^{N}\sum_{r=1}^{R}\sum_{i=1}^{I} I (y_{nri}=0) \log \{1 - F(S_{nri})\}+I (y_{nri}=1) \log \{F(S_{nri})\}
\end{split}
\label{ll_fun}
\end{equation}
where $F(S_{nri})$ being the probability of $Y_{nri}=1$, and $S_{nri}=\rho_r\sigma\theta'_n-\eta_r-\delta_i+\alpha$. As aforementioned, $F$ can be different link functions, among which, logit is the commonly used one. With the logit link function, Equation (\ref{ll_fun}) becomes 
\begin{equation} 
LL =- \sum_{n}\sum_{r}\sum_{i}  \log\{1+\exp(S_{nri})\} + \sum_{n}\sum_{r}\sum_{i} S_{nri}  I (y_{nri}=1).
\label{originalobjetive}
\end{equation}

Typically, the marginal maximum likelihood (MML) method \citep{bock1981marginal} is used to obtain the estimates for parameters $(\rho_r, \theta'_n, \delta_i, \eta_r, \alpha, \sigma), r=1,..., R, n=1,..., N, i=1,..., I$. To identify the parameters, the following constraints are imposed: $\sum_{i=1}^{I} \delta_i=0$, $\sum_{r=1}^{R}\eta_r/R=0$, $\sum_{n=1}^{N}\theta_n /N=0$, and $\sum_{n=1}^{N}\theta_n^2 /N=1$. Moreover, $\sigma$ is constrained to be greater than 0 because it is a scaling parameter and as discussed above, $\rho_r$ is constrained to be in a range of $0$ and $1$ to make the estimation more stable. In the MML, the student ability parameter $\theta_n$ is treated as random effects that follow $\mathcal{N}(0,1)$. Therefore, Equation \ref{originalobjetive}) is computed as
$$
 LL_{MAR} = \log \{\int ... \int \exp(LL) \phi(\theta_1)... \phi(\theta_N) d \theta_1 ...  d \theta_N \}=  \sum_{n} \log (M_n),
$$
 where
\begin{equation}
\begin{aligned}
M_n&=\int_{-\infty}^{+\infty} \prod_{r} \prod_{i}\{\mbox{F}(S_{nri})\}^{y_{nri}}  \{1-\mbox{F}(S_{nri})\}^{1-y_{nri} } \phi (\theta_n) d \theta_n.
\end{aligned}
\label{mml_integration_term}
\end{equation}
The likelihood evaluation in Equation (\ref{mml_integration_term}) involves integration $M_n$ for $n=1, 2, ..., N$. Due to a lack of analytic expressions for $M_n$, numerical approximations are needed. One possibility is to employ the Monte-Carol method to obtain the integral. Nevertheless, the method, involving computationally cost sampling, seems to be infeasible for educational assessments where large numbers of students and ratings are not rare.

To solve the difficulty in estimation, this study proposes the Laplace approximation method \citep{wolfinger1993laplace} for the integration process. Let  $h$ be the logarithm of the integrand in $M_n$, i.e., 
\begin{equation}
h(\theta_n)= \sum_{r \in J_{n} } \sum_{i=1}^{I} \{ I (y_{nri}=0) \log \left\{1- F(S_{nri})\right\}  
+ I (y_{nri}=1) \log F(S_{nri})\} - \frac{\theta_n^2}{2}. \label{eq:hierL}
\end{equation}
The integral $M_n$ can be represented as
$$
M_n=\frac{1}{\sqrt{2\pi}}\int_{-\infty}^{+\infty} \exp \left ( h(\theta_n) \right)  d \theta_n \approx
\frac{\exp \left (h(\theta_n^*)\right )}{\sqrt{|h''(\theta_n^*)|}},
$$
where $\theta_n^*=\arg \max_{\theta_n} h(\theta_n)$.
This leads to the Laplace approximation of the overall marginal log-likelihood function of $LL_{MAR}$ 
by
\begin{equation}
  LL_{MAR} \approx LL_{LAP} =  \sum_{n=1}^{n} \left[ h(\theta_n^*) -  0.5 \log( | h'' (\theta_n^*)|)  \right ].
 \label{laplace_app}
\end{equation}
For the GMF model with $F$ being a logistic function, we have
$$
 h'' (\theta_n^*)= - \sigma^2 \sum_{r \in J_{n} } \rho_r^2 \sum_{i=1}^{I}\frac{\exp\left(\textstyle \rho_r\sigma\theta_{n}^*-\eta_r-\delta_i+\alpha \right)}{\left\{1+ \exp\left(\textstyle \rho_r\sigma\theta_{n}^*-\eta_r-\delta_i+\alpha \right) \right\} ^2 }  -1.
$$
It should be noted that Equation (\ref{laplace_app}) is a function of all the other parameters,  while $\theta'_n$ is fixed at $\theta_n^*$, at which $g$ is maximized. This coincides with the HL estimation \citep{lee1996hierarchical}.  This estimation methodology bypasses the integration by maximizing the integrand directly by treating the integration variables as fixed parameters. This is for computation efficiency and has been proven to be asymptotically efficient. 
The HL function for parameter estimation is defined as $\sum_{n=1}^N h_n(\theta_n)$. Therefore, $\theta_n^*$ is also regarded as a valid estimator of $\theta'_n$. When $\theta_n^*$ is known,  we can then obtain all estimates except $\theta'_n$  by maximizing $LL_{LAP}$ given by Equation (\ref{laplace_app}). Hence, we can estimate $\theta'_n$ and others in different steps without integration. 

Specifically, to estimate the parameters in the GMF model (i.e., $\rho_r, \theta'_n, \eta_r, \delta_i, \sigma$, and $\alpha$), an iterative procedure proceeds as follows:
 \begin{itemize}
     \item Step 1:  Start with $\sigma=1$ and obtain initial estimates $ \hat{\theta'}_n, \hat{\eta}_r, \hat{\delta}_i$ and $\hat{\alpha}$ via `glm()' from R Package `stats';
     \item Step 2: Update $\hat{\theta'}_n$ (together with $\hat{\rho}_r$) for given 
     ($\hat{\eta}_r, \hat{\delta}_i$) (obtained in Step 1) via maximizing $h(\theta_n)$ in Equation (\ref{eq:hierL}) (assuming $\sigma=1$);
     \item Step 3: Update $\hat{\sigma}=\mbox{sd}(\hat{\theta'}_n)$ and $\theta_n^*=\hat{\theta}_n/\hat{\sigma}$, and then obtain $\hat{\rho}_r, \hat{\eta}_r, \hat{\delta}_i$ and $\hat{\alpha}$ by maximizing Equation (\ref{laplace_app});
     \item Step 4: Scale $\hat{\rho}_r=\hat{\rho}_r/ \max (\hat{\rho}_r)$ so that the maximum is 1, and update $\hat{\theta'}_n$ with $\hat{\eta}_r, \hat{\delta}_i$ and $\hat{\alpha}$ by maximizing the HL, $ h_n(\theta'_n)$;
     \item Step 5: Repeat Steps 3 and 4 until the procedure reaches the predefined convergence criteria, as in, the maximum number of iterations (10 in this work) or the change of scale estimates smaller than 0.01.    
 \end{itemize}

With the estimates for the parameters in the GMF model, the overall capability for each rater can be calculated:
\begin{eqnarray}
\bar{\kappa}_r^{GMF}& =&  \Delta^{-1}\sigma \rho_r\int_{-\infty}^{+\infty} \mu_{nr}(1-\mu_{nr}) \phi(\theta) d \theta  \nonumber  \\
&=& 
\Delta^{-1} \int_{-\infty}^{+\infty} \frac{\displaystyle\exp(x)}{\displaystyle[1+\exp(x)]^2} \phi(\frac{\displaystyle x -\eta_r}{\displaystyle \rho_r\sigma}) dx \nonumber \\ 
&\approx& 4 \rho_r \sqrt{\frac{\displaystyle 2+\sigma^2}{\displaystyle 2}} \frac{\displaystyle \exp \left \{x^*-\frac{1}{2} \left (\frac{x^*-\eta_r}{\rho_r \sigma} \right)^2 \right \} }{\displaystyle (1+\exp(x^*))\sqrt{(1+\exp(x^*))^2+2\rho_r^2\sigma^2 \exp(x^*)}},
\label{kappa_formula}
\end{eqnarray}
where $x^*$ is the $x$ value that maximizes the integrand and is calculated by setting the derivative to 0, leading to
$$
x^*= \rho_r^2\sigma^2\frac{1-\exp (x^*)}{1+\exp(x^*)}+\eta_r.
$$

With respect to the uncertainties associated with the $\bar{\kappa}_r^{GMF}$, the variance can be obtained using the Delta method as
\begin{equation}
\left ( \frac{\partial \bar{\kappa}_r^{GMF} (\zeta_r)}{\partial \zeta_r} \right ) Var(\zeta_r) \left(\frac{\partial \bar{\kappa}_r^{GMF} (\zeta_r)}{\partial \zeta_r} \right)^T,
\end{equation}
where $\zeta_r=(\hat{\sigma},\hat{\rho}_r,\hat{\eta}_r)$.

\section{Simulation Studies}\label{numerical-simulation}

The simulation study aimed to address four distinct research questions. First, how well can the proposed estimation method use the HL and Laplacian approximation to estimate the parameters in the GMF model? Second, how does the proposed capability index perform in measuring raters' accuracy? third, what are the consequences when the fitting model was misspecified for the data? and fourth, how to improve raters' capability?

To address these research questions, two simulation studies were conducted. Study 1 was designed to provide insights into the first three research questions, while Study 2 focused on the second and fourth questions. Each condition was subjected to two hundred replications to ensure robust findings.

\subsection{Simulation Study 1}

In the first simulation study, we maintained a constant number of raters (20), students (50), and items (40), with each rater evaluating all students and items. The responses were generated using the GMF model, wherein the discrimination parameter ($\rho_r$) for the $r$-th rater was assigned a value of $r/20$, and the severity parameter (i.e., $\eta_r$) was set to $(21-r)/9-1$ for $r=1,...,20$. Consequently, rater 1 embodies characteristics of low discrimination and high severity ($\rho_1 = 0.05$, $\eta_1 = 2.22$), while rater 20 is marked by high discrimination and minimal bias ($\rho_{20} = 1$, $\eta_{20} = 0.11$). This design anticipates that the capability index values of rater 20 will surpass those of rater 1. Moreover, student ability (i.e., $\theta_n$) was generated from a normal distribution $\mathcal{N}(0,0.5^2)$, item difficulty (i.e., $\delta_i$) followed a pattern of $i/(20-1)$ for $i=1,...,40$, and the intercept parameter was fixed at 0.5.

The generated data were used to fit the GMF model and the TFM utilizing the proposed estimation method. The calculation of $\bar{\kappa}_r$ was carried out based on Equation (\ref{kappa_formula}). It's important to mention that under the TFM, $\rho$ is universally assumed to be 1 for all raters as per the definition. Consequently, when computing $\bar{\kappa}_r$ for the TFM, the value of $\rho_r$ was set to 1.

The study evaluated the bias and root mean square error (RMSE) as the dependent variables for key parameters, including the discrimination parameter ($\rho_r$), severity parameter ($\eta_r$), overall capability parameter ($\bar{\kappa}_r$), individual student ability ($\theta_n$), and scale parameter ($\sigma$) under both fitted models. For instance, with regard to $\rho_r$, the bias and RMSE were defined as follows:
\begin{equation} \label{bias}
\text{Bias}(\hat{\rho}_r)=\sum{r=1}^{L}(\hat{\rho}_{lr}-\rho_r)/L,
\end{equation}
and
\begin{equation} \label{rmse}
\text{RMSE}(\hat{\rho}_r)=\sqrt[]{\sum{l=1}^{L}(\hat{\rho}_{lr} - \bar{\hat{\rho}}_r)^2/L},
\end{equation}
where $\rho_r$ represented the generated value of $\rho$ for rater $r$, $\hat{\rho}_{lr}$ stood for the estimation for rater $r$ in the $l$-th replication, and $\bar{\hat{\rho}}_r$ indicated the average estimate for rater $r$ across $L$ replications.

\begin{figure}[!ht]
    \centering
    \includegraphics[width=\textwidth]{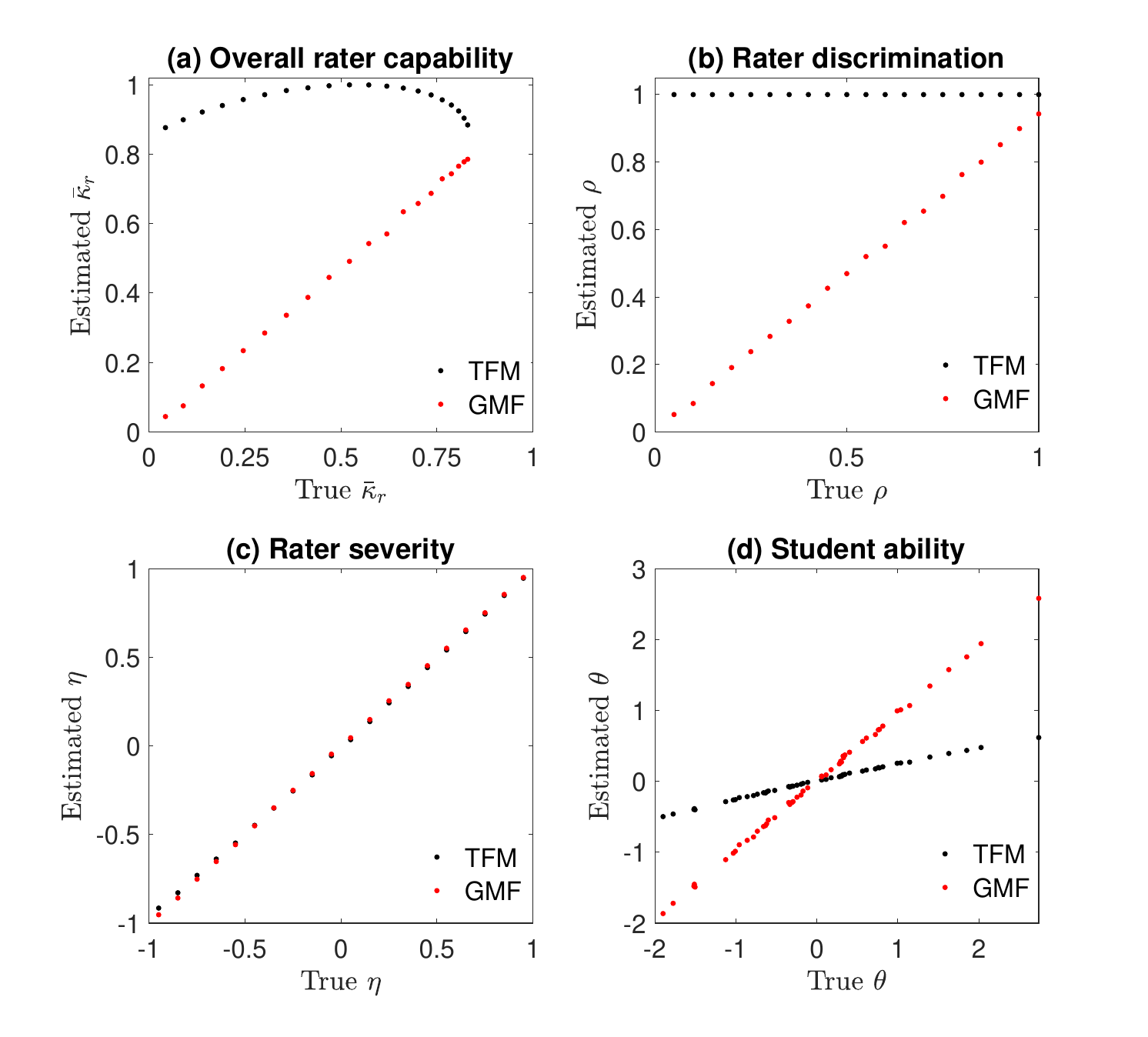}
    \caption{Results of parameter estimates in simulation study 1. }
    \label{s11curve}
\end{figure}

The outcomes are visually depicted in Figure~\ref{s11curve} and quantitatively detailed in Table~\ref{s11numeric} for rater parameters, Table~\ref{s11numeric_abi} for student ability, and Appendix E's Table~\ref{s11numeric_delta} for item difficulty. Several noteworthy insights emerge as outlined below.

To begin, the introduced estimation approach effectively produces precise estimates for rater, student, and item parameters. This is evident in Figure~\ref{s11curve}, which illustrates the alignment of estimates with true parameter values under the GMF model. The estimates nearly adhere to the main diagonals, signifying their high accuracy.

Next, the newly proposed index effectively captures raters' capability within the GMF model. Subplot (a) in Figure~\ref{s11curve} presents the ordered arrangement of dots (representing raters) based on their assigned numbers. In this arrangement, rater 1 displays the smallest $\bar{\kappa}$ value (0.04), while rater 20 showcases the highest value (0.79). These outcomes directly align with our anticipations, clearly indicating that rater 1 possesses the least capability, whereas rater 20 exhibits the highest capability.

Thirdly, when fitting the TFM, the estimates of rater severity remained unbiased, aligning with our expectations. However, an inconsistency emerged in the estimates of student ability. Specifically, estimates for high-ability students were lower than anticipated, while estimates for low-ability students were higher, indicating a compression of the $\theta$ scale. This occurrence can be attributed to the data generation process, which utilized the GMF model where the rater discrimination ($\rho$) and student ability ($\theta'$) were multiplied using true $\rho$ values ranging from 0.05 to 1.00. In contrast, the TFM imposed a fixed value of 1 for $\rho$, leading to the observed compression of $\theta$ estimates.

Lastly, it was observed that the overall capability $\bar{\kappa}_r$ tended to be overestimated when using the TFM. This outcome is linked to the fixed value of $\rho$ at 1 within the TFM. The discrepancy arises from the fact that fixing $\rho$ results in an overestimation of the discrimination parameters within the TFM, ultimately leading to an overestimation of the $\bar{\kappa}_r$ values.

\begin{table}[!ht]
\centering
\footnotesize
\caption{Bias and root mean square error (RMSE) for estimates of rater parameters in simulation study 1} 
\begin{tabular}{lcccccccccccccc}
\hline
     &\multicolumn{4}{c}{$\hat{\bar{\kappa}}_r$}&&\multicolumn{4}{c}{$\hat{\rho}_r$}&&\multicolumn{4}{c}{$\hat{\eta}_r$}\\
     \cmidrule(r){2-5}  \cmidrule(r){7-10} \cmidrule(r){12-15} 
     &\multicolumn{2}{c}{TFM}& \multicolumn{2}{c}{GFM}& &\multicolumn{2}{c}{TFM}& \multicolumn{2}{c}{GFM}& &\multicolumn{2}{c}{TFM}& \multicolumn{2}{c}{GFM}\\     
\cmidrule(r){2-3} \cmidrule(r){4-5}  \cmidrule(r){7-8} \cmidrule(r){9-10} \cmidrule(r){12-13} \cmidrule(r){14-15} 
Rater & Bias & RMSE& Bias & RMSE&  & Bias & RMSE&  Bias & RMSE& & Bias & RMSE&  Bias & RMSE\\
\hline
01&	0.83&	0.83&	 0.00&	0.08&	&	0.95&	0.95&	 0.00&	0.09&	&	0.00&	0.05&	0.00&	0.05\\
02&	0.81&	0.81&	-0.01&	0.08&	&	0.90&	0.90&	-0.02&	0.09&	&	0.00&	0.04&	0.01&	0.04\\
03& 0.78&	0.78&	-0.01&	0.09&	&	0.85&	0.85&	-0.01&	0.10&	&	0.00&	0.05&	0.00&	0.05\\
04& 0.75&	0.75&	-0.01&	0.09&	&	0.80&	0.80&	-0.01&	0.09&	&	0.00&	0.05&	0.01&	0.05\\
05& 0.71&	0.71&	-0.01&	0.09&	&	0.75&	0.75&	-0.01&	0.09&	&	-0.01&	0.05&	0.00&	0.05\\
06& 0.67&	0.67&	-0.02&	0.09&	&	0.70&	0.70&	-0.02&	0.09&	&	-0.01&	0.04&	0.00&	0.04\\
07& 0.63&	0.63&	-0.02&	0.10&	&	0.65&	0.65&	-0.02&	0.10&	&	-0.01&	0.05&	0.00&	0.05\\
08& 0.58&	0.58&	-0.03&	0.09&	&	0.60&	0.60&	-0.03&	0.09&	&	-0.01&	0.05&	0.01&	0.05\\
09& 0.53&	0.53&	-0.02&	0.10&	&	0.55&	0.55&	-0.02&	0.10&	&	-0.01&	0.04&	0.00&	0.04\\
10& 0.48&	0.48&	-0.03&	0.12&	&	0.50&	0.50&	-0.03&	0.12&	&	-0.01&	0.05&	0.00&	0.05\\
11& 0.43&	0.43&	-0.03&	0.11&	&	0.45&	0.45&	-0.03&	0.11&	&	0.00&	0.05&	0.00&	0.05\\
12& 0.38&	0.38&	-0.05&	0.12&	&	0.40&	0.40&	-0.05&	0.12&	&	-0.01&	0.05&	-0.01&	0.05\\
13& 0.28&	0.28&	-0.04&	0.11&	&	0.30&	0.30&	-0.05&	0.11&	&	0.00&	0.05&	0.00&	0.05\\
15& 0.24&	0.24&	-0.05&	0.11&	&	0.25&	0.25&	-0.05&	0.12&	&	0.00&	0.05&	0.00&	0.05\\
16& 0.19&	0.19&	-0.03&	0.11&	&	0.20&	0.20&	-0.04&	0.13&	&	0.00&	0.05&	-0.01&	0.05\\
17& 0.15&	0.15&	-0.04&	0.10&	&	0.15&	0.15&	-0.05&	0.12&	&	0.01&	0.05&	0.00&	0.05\\
18& 0.12&	0.12&	-0.04&	0.10&	&	0.10&	0.10&	-0.05&	0.12&	&	0.02&	0.06&	0.00&	0.06\\
19& 0.08&	0.08&	-0.04&	0.08&	&	0.05&	0.05&	-0.05&	0.10&	&	0.02&	0.06&	-0.01&	0.06\\
20& 0.05&	0.06&	-0.04&	0.08&	&	0.00&	0.00&	-0.06&	0.10&	&	0.04&	0.06&	0.00&	0.06\\
\hline
\end{tabular}
\label{s11numeric}
\begin{tablenotes}
\item \textit{Note}: $\hat{\bar{\kappa}}_r$ = estimates of overall capability index for rater $r$; $\hat{\rho}_r$ = estimates of rater discrimination for rater $r$; $\hat{\eta}_r$ = estimates of rater severity for rater $r$; TFM = three-facets model; GMF = generalized multi-facets model. The bias and RMSE of scale parameter $\sigma$ estimate are 0.02 and 0.05.
\end{tablenotes}
\end{table}

\begin{table}[!ht]
\centering
\footnotesize
\caption{Bias and root mean square error (RMSE) for estimates of student ability in simulation study 1} 
\begin{tabularx}{\textwidth}{cccccccccccc}
\hline
&&\multicolumn{4}{c}{Top 25 students}& &&\multicolumn{4}{c}{Bottom 25 students}\\  
\cmidrule(r){3-6}  \cmidrule(r){9-12} 
&&\multicolumn{2}{c}{TFM}& \multicolumn{2}{c}{GFM}& &&\multicolumn{2}{c}{TFM}& \multicolumn{2}{c}{GFM}\\     
\cmidrule(r){3-4} \cmidrule(r){5-6}  \cmidrule(r){9-10} \cmidrule(r){11-12} 
Student& $\theta_n$ & Bias & RMSE& Bias & RMSE& Student&  $\theta_n$ & Bias & RMSE&  Bias & RMSE\\
\hline
28&	2.74&	-2.12&	2.12&	-0.15&	0.32&	01&	-0.11&	0.10&	0.12&	0.02&	0.24 \\
37&	2.03&	-1.55&	1.55&	-0.08&	0.28&	50&	-0.17&	0.14&	0.16&	0.04&	0.24 \\
38&	1.85&	-1.41&	1.41&	-0.09&	0.32&	09&	-0.20&	0.16&	0.17&	0.00&	0.25 \\
12&	1.63&	-1.23&	1.24&	-0.05&	0.26&	35&	-0.25&	0.19&	0.21&	0.02&	0.25 \\
40&	1.40&	-1.05&	1.05&	-0.05&	0.28&	26&	-0.29&	0.23&	0.24&	0.01&	0.27 \\
10&	1.15&	-0.87&	0.88&	-0.07&	0.30&	45&	-0.31&	0.24&	0.25&	0.01&	0.24 \\
04&	1.04&	-0.77&	0.78&	-0.02&	0.29&	41&	-0.33&	0.25&	0.27&	0.01&	0.27 \\
20&	0.99&	-0.73&	0.74&	0.00&	0.27&	21&	-0.35&	0.28&	0.28&	0.05&	0.26 \\
39&	0.81&	-0.61&	0.61&	-0.03&	0.27&	27&	-0.52&	0.39&	0.40&	0.01&	0.26 \\
19&	0.77&	-0.58&	0.58&	-0.04&	0.27&	49&	-0.60&	0.46&	0.47&	0.05&	0.26 \\
11&	0.76&	-0.56&	0.57&	-0.03&	0.27&	43&	-0.62&	0.47&	0.48&	0.02&	0.27 \\
33&	0.72&	-0.54&	0.55&	-0.06&	0.28&	44&	-0.63&	0.47&	0.48&	0.01&	0.26 \\
22&	0.61&	-0.45&	0.46&	0.00&	0.27&	06&	-0.66&	0.50&	0.50&	0.02&	0.25 \\
29&	0.56&	-0.42&	0.42&	0.00&	0.26&	34&	-0.73&	0.55&	0.56&	0.03&	0.24 \\
32&	0.40&	-0.28&	0.29&	0.01&	0.27&	16&	-0.78&	0.58&	0.59&	0.00&	0.27 \\
46&	0.35&	-0.25&	0.26&	0.03&	0.31&	08&	-0.86&	0.65&	0.65&	0.03&	0.25 \\
17&	0.33&	-0.24&	0.25&	0.01&	0.26&	05&	-0.96&	0.73&	0.73&	0.06&	0.25 \\
48&	0.32&	-0.23&	0.24&	0.03&	0.29&	02&	-1.01&	0.75&	0.75&	0.02&	0.25 \\
24&	0.30&	-0.23&	0.24&	-0.03&	0.25&	14&	-1.03&	0.77&	0.77&	0.02&	0.23 \\
30&	0.29&	-0.22&	0.23&	-0.01&	0.26&	36&	-1.13&	0.84&	0.84&	0.02&	0.25 \\
47&	0.28&	-0.21&	0.23&	-0.03&	0.27&	07&	-1.50&	1.10&	1.11&	0.01&	0.24 \\
18&	0.18&	-0.12&	0.15&	-0.01&	0.28&	23&	-1.51&	1.13&	1.13&	0.06&	0.23 \\
15&	0.11&	-0.08&	0.12&	-0.02&	0.28&	13&	-1.52&	1.12&	1.13&	0.04&	0.26 \\
25&	0.06&	-0.03&	0.08&	0.02&	0.23&	42&	-1.77&	1.31&	1.32&	0.05&	0.24 \\
31&	0.06&	-0.04&	0.08&	0.01&	0.23&	03&	-1.90&	1.40&	1.40&	0.03&	0.24 \\
\hline
\end{tabularx}
\label{s11numeric_abi}
\begin{tablenotes}
\item \textit{Note}: $\theta_n$ = true value of ability for student $n$; TFM = three-facets model; GMF = generalized multi-facets model. Due to space constraints, only the estimates for the top 25 and the bottom 25 students are reported in this table.
\end{tablenotes}
\end{table}

\subsection{Simulation Study 2}

Simulation study 2 aimed to demonstrate the enhancement of overall capability through the modification of raters' severity. The simulation design mirrored the empirical study, wherein each of the four raters assessed a subset of students across four writing topics. True values from the estimates of the empirical study were adopted for data generation within the GMF model. However, the parameter $\eta$ was varied within a range of -2.5 to 2.5, with increments of 0.1. Subsequently, the generating model (i.e., GMF) was applied to the generated data.

\begin{figure}[htp]
  \centering
    \includegraphics[width=0.9\textwidth]{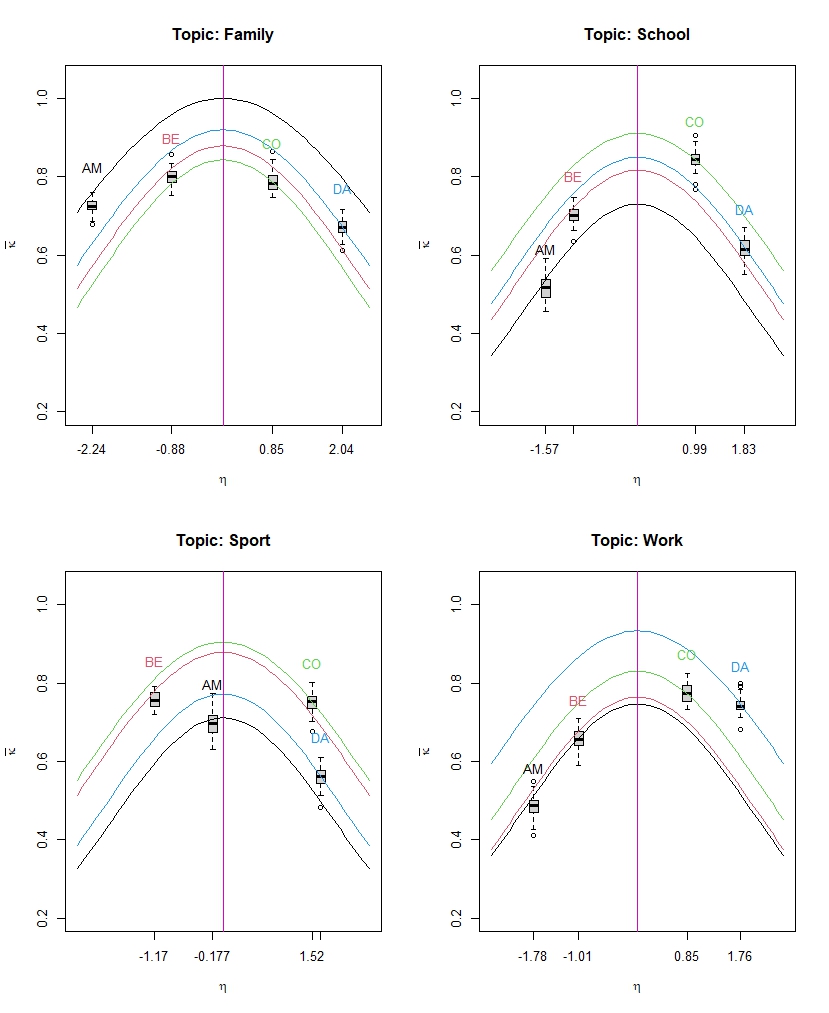}
    \caption{The overall capability index curves in simulation study 2}    
    \label{kappa_implication}
\end{figure}

The outcomes are graphically depicted in Figure~\ref{kappa_implication}, illustrating the estimated overall capability curves associated with distinct severity values. On the x-axis, the actual severity values for each rater are displayed, while the boxes portray the estimates obtained across various replications. Figure~\ref{kappa_implication} holds noteworthy implications, enabling the identification of a rater's capability level and offering insights into enhancing their capability by fine-tuning their severity levels.

Using rater AM as an illustration, the results demonstrate that the rater exhibits a high level of leniency when evaluating the topics of family, school, and work while displaying a relatively unbiased approach when rating the topic of sport. Of greater significance is the revelation that the rater's overall capability can be enhanced by elevating their severity in rating the aforementioned three topics. These insights hold the potential to facilitate the oversight of rater performances and offer tailored feedback to enhance their rating accuracy.

\section{An Empirical Study}

The proposed capability index was subjected to evaluation through its application to a dataset sourced from the R package Tam~\citep{tam2022}. This dataset encompassed 1,452 instances involving 363 students, with ratings assigned by four raters (referred to as ``AM'', ``BE'', ``CO'', and ``DA'') across four distinct writing subjects (``family'', ``school'', ``sport'', and ``work''), and five specific criteria (``specificity'', ``coherence'', ``structure'', ``grammar'', ``content''). Our investigation was driven by two central inquiries: 1. What was the overall capability exhibited by the raters, both in a generalized context and when rating distinct topics? and 2. What were the approximations for raters' attributes in terms of discrimination and severity?

The initial four-category ratings were transformed into a dichotomous format, where a score of $0, 1, 2$ was consolidated as 0, and a score of $3$ was denoted as 1. The distribution of ratings assigned by each rater across various topics spanned a range of $365$ to $535$, with cumulative scores varying from $103$ to $392$, as outlined in Table~\ref{emp_res}. Employing the Generalized Multi-Facets (GMF) model, the dataset was analyzed utilizing the proposed estimation technique, leading to the computation of the comprehensive capability index.

The estimated item difficulties for the criteria of specificity, coherence, structure, grammar, and content are as follows: $-1.54$, $-1.45$, $0.19$, $0.76$, and $2.04$, respectively. Consequently, it can be deduced that writing specificity is the least challenging aspect, whereas content poses the most significant difficulty for the students. In terms of student ability, the estimated value is $\hat{\sigma}=2.51$.

The approximated values for rater discrimination, rater severity, and the comprehensive rater capability are displayed in Table~\ref{emp_res}. The mean overall rater capability is computed for various topics: family (0.76), school (0.70), sport (0.64), and work (0.68). This analysis suggests that, on average, raters exhibit higher capability in evaluating writings on the subject of family, while their competence is relatively lower when assessing writings centered around the topic of sport. These findings underscore the significance of providing raters with appropriate training to effectively evaluate writings on the subject of sport.

The findings also unveil variations in raters' capabilities across different topics. For instance, rater AM exhibited higher capability ($\bar{\kappa}=0.76$) when evaluating essays on the subject of family, yet displayed lower capability when rating essays on the remaining three topics. In contrast, rater DA demonstrated more proficient ratings for essays related to the topic of work ($\bar{\kappa}=0.74$). Both AM and DA exhibited notable limitations when assessing essays on the topic of sport. Conversely, raters BE and CO seemed proficient in rating diverse topics, with CO displaying notable competence in evaluating essays about the subject of school.

To validate the outcomes, we extracted two distinct sets of ratings, as in, the ratings on the five items provided by rater AM for the topic of sport, which displayed the lowest $\bar{\kappa}$ value of 0.48. and the ratings by rater CO for the topic of school, showcasing the highest $\bar{\kappa}$ value of 0.90. Additionally, we obtained the corresponding estimates of student ability for these two sets of ratings and computed the point-biserial correlation between each rating and the ability estimates. The outcomes revealed correlations of 0.34, 0.40, 0.59, 0.38, and 0.49 for rater AM, and correlations of 0.56, 0.65, 0.58, 0.60, and 0.52 for rater CO. This suggests that CO demonstrates greater capability in assigning a score of 1 to students with higher abilities and a score of 0 to those with lower abilities compared to AM.

\begin{table}[!ht]
\centering
\small
\caption{Number of ratings, summed scores, and estimates of rater parameters for different topics with generalized multi-facets model} 
\begin{tabular}{llccrrr}
\hline
Rater&Topic& Number of ratings &Summed scores&$\hat{\rho}$&$\hat{\eta}$&$\hat{\bar{\kappa}}$\\
\hline
\multirow{4}{*}{AM}&Family&  440&338 &1.00&	-2.24&	0.76\\
&School	&525&392 &0.50&	-1.57&	0.54\\
&Sport	&370&286  &0.48&	-1.77&	0.48\\
&Work	&455&335  &0.52&	-1.78&	0.51\\
\cline{1-7}
\multirow{4}{*}{BE}&Family	&435 &273 &0.71&	-0.88&	0.82\\
&School	&470 &298 &0.61&	-1.09&	0.72\\
&Sport	&535&320  &0.71&	-1.17&	0.78\\
&Work	&380&233  & 0.54&	-1.01&	0.68\\
\cline{1-7}
\multirow{4}{*}{CO}&Family	&450&195 &0.65&	0.85&	0.78\\
&School	&535&220 &0.88&	0.99&	0.90\\
&Sport	&485&178 &0.76&	1.52&	0.75\\
&Work	&440&183 &0.63&	0.85&	0.77\\
\cline{1-7}
\multirow{4}{*}{DA}&Family	&365&103 &0.79&	2.04&	0.67\\
&School	&520&155 &0.66&	1.83&	0.62\\
&Sport	&460&142 &0.55&	1.67&	0.56\\
&Work	&395&106 &0.82&	1.76&	0.74\\
\hline
\end{tabular}
\label{emp_res}
\end{table}

A more in-depth analysis of the outcomes exposes the degrees of discrimination and severity exhibited by raters across various topics. Notably, AM tends to be an overly lenient rater, whereas DA demonstrates to be a severe rater. Additionally, these two raters possess relatively low discrimination parameters. The combination of their discrimination and severity levels consequently contributes to their lower capability. Recall that simulation study 2 demonstrated that raters can adjust their severity levels to enhance their overall capability. Given these findings, it is reasonable to anticipate that AM could potentially enhance their capability by increasing their severity, while DA might improve their capability by decreasing their severity.

\section{Conclusion and Discussion}

In this study, we introduce a novel single-value measure designed to measure the capability of individual raters. This measure draws upon the partial derivatives of a rater's passing rate about the student's ability. The process of calculating this index under various IRT models is underpinned by several mathematical proofs. To enhance estimation efficiency, we also propose a method for parameter estimation using the HL and Laplacian approximation, specifically tailored for the popular GMF model. The approximated likelihood makes it possible to efficiently evaluate and handle large educational datasets that contain numerous students or raters without the need for integration. The outcomes of simulation studies underscore the accuracy of the proposed approximate likelihood estimation method in parameter estimation.

Through comprehensive simulation studies, we examine the capability index to accurately assess raters' accuracy, identify key rater attributes influencing their capability, and explore methods to enhance their performance. The findings indicate that the proposed index effectively captures raters' capability under the GMF model. Both rater discrimination and severity contribute to their overall capability level. Consequently, when fitting the TFM, where only the severity parameter can be reliably estimated, the resultant capability estimates may be inadequate. Given that the GMF model includes the TFM as a special case and its estimation procedure is readily available, we recommend utilizing the GMF model for real-world data analysis when applying the capability index. Moreover, the results highlight the potential for enhancing raters' capability through adjustments to their severity. This implies that the proposed capability index could be viably employed for monitoring raters' assessments and providing tailored feedback to improve their rating accuracy.

We demonstrate the practical viability of the new measure by applying it to a real dataset in conjunction with the proposed estimation method. Besides item difficulty, student ability, and rater discrimination and severity estimates, the results reveal the heterogeneity of raters' capabilities in general and in different topics. 

Despite the encouraging outcomes, this study does have its limitations. Firstly, the flexibility of $\mu_{nri}$ permits the incorporation of various variable combinations. Thus, in future research, it could be formulated as a function of individual rater variables (e.g., experience, year) to encompass the diversity in their capabilities. Secondly, further simulation studies are necessary to thoroughly evaluate the effectiveness of the proposed capability index in different realistic scenarios. For instance, a simulation setting that mirrors real-world conditions where certain ratings, or a portion of them, are influenced by noise could be considered. Moreover, while the current simulation studies utilized a complete design where each rater assessed all students and items in Study 1 and an incomplete and unbalanced design where raters evaluated a specific number of students in Study 2, additional research is required where the linkage sets are manipulated, as shown in \citet{casabianca2023using}.

Lastly, due to time constraints, this study could not extensively explore the newly introduced index within the probit model and the HRM. However, given that this study has already provided the computation and mathematical evidence, it is straightforward for future research to implement the index in these models. Moreover, it's important to emphasize that our suggested capability index is grounded in the ability of test-takers. Therefore, if an IRT model can precisely estimate all the relevant parameters, the index can offer a more dependable assessment of rater capability. To illustrate, in cases where certain raters exhibit inappropriate behavior, we will opt for a robust IRT model instead of the conventional one to derive the parameters required for computing the capability index.

\vspace{\fill}\pagebreak

\vspace{12pt}
\noindent \textbf{\large {Declaration of interest}}

All the authors declare that they have no competing interests.

\bibliographystyle{unsrtnat}


\clearpage
\newpage
\appendix
\setcounter{table}{0} \renewcommand{\thetable}{A.\arabic{table}}
\renewcommand\theHtable{Appendix.\thetable}
\setcounter{figure}{0} \renewcommand{\thefigure}{A.\arabic{figure}}
\renewcommand\theHfigure{Appendix.\thefigure}

\section*{Appendix A: $\Delta$ in the Three-facets Model}

We proceed to demonstrate the following equality:
$$
\Delta =\sup_r \int_{-\infty}^{+\infty} \frac{\exp(x)}{[1+\exp( x)]^2} \phi(x-\delta_i-\eta_r) dx = \int_{-\infty}^{+\infty} 
 \frac{\exp(x)}{[1+\exp( x)]^2} \phi(x) dx.
$$
Let us first define the function $T(z)$ as follows:
$$
T(z)=\int_{-\infty}^{+\infty} \frac{\exp(x)}{(1+\exp(x))^2} \phi (x+z)dx,
$$
where $z=-\eta_r$. Now, we can compute the derivative of $T(z)$ with respect to $z$:
$$
T'(z) = - \int_{-\infty}^{\infty} (x+z)\frac{\exp(x)}{(1+\exp(x))^2} \phi (x+z)dx = -z
\int_{-\infty}^{\infty} \frac{\exp(x)}{(1+\exp(x))^2} \phi (x+z)dx ,
$$
It is evident that $T'(z) < 0$ for $z>0$ and $T'(z) > 0$ for $z<0$. This indicates that $T(z) < T(0)$ for any $z\neq 0$. Hence ,we can express $\Delta$ as follows: 
$$
 \Delta = \sup_r \int_{-\infty}^{+\infty} 
 \frac{\exp(x)}{[1+\exp( x)]^2} \phi(x-\eta_r) dx = 
 \sup_r T(\eta_r)
=T(0) = \int_{-\infty}^{+\infty} 
 \frac{\exp(x)}{[1+\exp( x)]^2} \phi(x) dx,
$$
In other words, the supremum (i.e., $\Delta$) is obtained when $z=0$ or $\eta_r=0$.

\section*{Appendix B: $\Delta  $ in the Generalized Multi-facets Model}

We proceed to demonstrate the following equality:
$$
 \Delta = \sigma \sup_r   \rho_r \int_{-\infty}^{+\infty} 
 \frac{\exp(\rho_r\sigma\theta-\eta_r)}{[1+\exp(\rho_r\sigma\theta-\eta_r)]^2} \phi(\theta) d\theta =  \int_{-\infty}^{+\infty} 
 \frac{\exp(x)}{[1+\exp( x)]^2} \phi(x/\sigma) dx.
$$
Let us introduce the variables $z=\eta_r$ and $x=\rho_r\sigma\theta-z$. We define a function $T(z,\rho_r)$:
$$
T(z, \rho_r)= \sigma
\rho_r \int_{-\infty}^{+\infty} 
 \frac{\exp(\rho_r\sigma\theta-\eta_r)}{[1+\exp(\rho_r\sigma\theta-\eta_r)]^2} \phi(\theta) d\theta =
\int_{-\infty}^{+\infty} \frac{\exp (x)}{[1+ \exp(x)]^2} \phi (\frac{x+z}{\rho_r \sigma}) dx.
$$
Let's calculate the partial derivative of $T(z,\rho_r)$ with respect to $z$, denoted as $G(z,\rho_r)$:
$$
G(z,\rho_r)=\frac{\partial T(z,\rho_r) }{\partial z}= - \int_{-\infty}^{+\infty} \frac{x+z}{\rho_r \sigma}\frac{\exp (x)}{[1+ \exp(x)]^2} \phi (\frac{x+z}{\rho_r \sigma}) dx=-\frac{z}{\rho_r \sigma}\int_{-\infty}^{+\infty} \frac{\exp (x)}{[1+ \exp(x)]^2} \phi (\frac{x+z}{\rho_r \sigma}) dx.
$$
We observe that $G(z,\rho_r) < 0$ for $z>0$ and $G(z,\rho_r) > 0$ for $z<0$. This means $T(z,\rho_r)$ is decreasing with respect to $|z|$, which implies $T(|z|,\rho_r) < T(0, \rho_r)$ for any $z\neq 0$. 
Furthermore, let's examine the derivative of $T(0,\rho_r)$ with respect to $\rho_r$, denoted as $g(0,\rho_r)$:
$$
g(0,\rho_r)=\frac{\partial T(0,\rho_r) }{\partial \rho_r}=\frac{1}{\sigma^2 \rho_r^3}\int_{-\infty}^{+\infty} x^2 \frac{\exp (x)}{[1+ \exp(x)]^2} \phi (\frac{x}{\rho_r \sigma}) dx \geq 0.
$$
This indicates that $g(0,\rho_r)>0$ for for any $\rho_r >0$. Thus, $T(0,\rho_r)$ is a monotonically increasing function with respect to $\rho_r$. Therefore, $T(0,\rho_r)$ reaches its maximum when $\rho_r$ reaches its maximum bound 1, which leads to
\begin{equation*}
 \Delta = \sup_{r} T(z, \rho_r) = \max_{\rho_r} T(0, \rho_r)  =  \int_{-\infty}^{+\infty} 
 \frac{\exp(x)}{[1+\exp( x)]^2} \phi(x/\sigma) dx.
\end{equation*}

\section*{Appendix C: $\Delta$ in Probit Models}

We proceed to demonstrate the following equality:
$$
\Delta =\sup_r \int_{-\infty}^{+\infty} \frac{\phi(S_{nri})}{\sigma_r} \phi(\theta) d \theta=\frac{1}{\sqrt{2\pi}}.
$$

According to the definition of $\Delta$, 
 \begin{eqnarray*} 
   \Delta  & = &  \sup_r \int_{-\infty}^{+\infty} \frac{\phi(S_{nri})}{\sigma_r} \phi(\theta) d \theta\\ 
  & = &  \frac{1}{\sqrt{2\pi}} \sup_r  \frac{1}{\sigma_r}\int_{-\infty}^{+\infty}  
   \exp(- S_{nri}^2/2)\phi(\theta) d \theta \\ 
   & = &  \frac{1}{\sqrt{2\pi}} \sup_r \frac{1}{\sqrt{1+\sigma_r^2}} 
   \exp\left\{-\frac{\sigma_r^2\eta_r^2}{2\sqrt{1+\sigma_r^2}}\right\} \\ 
   & = &  \frac{1}{\sqrt{2\pi}}. 
\end{eqnarray*}
   The last equation holds because the maximum value is reached when $\eta_r=0$ and $\sigma_r=0$.

\section*{Appendix D: $\Delta$ in the Hierarchical Rater Model}
As mentioned in the main text, the HRM consists of two levels. For illustration, let the first level be a latent class signal detection theory model:
\begin{equation} \label{ap_hrm_l1} 
P(Y_{nri}=1)= \textit{F} \left(  c_{r}   - a_{ri} \xi_{ni} \right), 
\end{equation}
and  at the second level is the Rasch model, 
\begin{equation} \label{ap_hrm_l2} 
P(\xi_{ni}=1)= \frac{\exp(\theta_n-\delta_i+\alpha)}{ 1+\exp(\theta_n-\delta_i+\alpha)},
\end{equation}
where the mean $\theta_n$, again, is assumed to be 0, and the intercept  $\alpha$ is to control the overall marginal passing rate. The parameters are defined as in Equation (\ref{hrm_l1}) in the main text. In calculating the $\kappa$ functions, we assume $\delta_i=0$. Based on Equations (\ref{ap_hrm_l1}) and  (\ref{ap_hrm_l2}), we have $F_{2,1}=F(c_{ri}-a_{ri})$, $F_{2,0}=F(c_{ri})$, $F_0=\frac{1}{1+\exp(\theta_n+\alpha)}$, and $F_1=\frac{\exp(\theta_n+\alpha)}{1+\exp(\theta_n+\alpha)}$.

For this model, we can derive the expression for $\frac{\partial\mu_{nri}}{\partial \theta}$ as follows:
\begin{eqnarray} \label{ap_hrm_4} 
\frac{\partial \mu_{nri}}{\partial\theta}  &= &\frac{\partial (\sum_{k=0}^{1} F_k  F_{2,k}) }{\partial\theta}\\ \nonumber
&= &\frac{\partial} {{\partial\theta}} \left[\frac{1}{1+\exp(\theta_n+\alpha)}  F(c_{ri}) \right] + \frac{\partial} {{\partial\theta}} \left[ \frac{\exp(\theta_n+\alpha)}{1+\exp(\theta_n+\alpha)} F(c_{ri}-a_{ri}) \right] \\ \nonumber
& = &\frac{\partial} {{\partial\theta}} \left[\frac{1}{1+\exp(\theta_n+\alpha)} \right] F(c_{ri}) + \frac{\partial} {{\partial\theta}} \left[\frac{\exp(\theta_n+\alpha)}{1+\exp(\theta_n+\alpha)} \right] F(c_{ri}-a_{ri}) \\ \nonumber
& = &F_0 F_1 \left(F_{2,1} -F_{2,0}\right).
\end{eqnarray}
 
The supremum of Equation (\ref{ap_hrm_4}) is reached when $F_{2,1} -F_{2,0}=1$, and the integration part is a constant independent of the raters. Therefore, 
$$
\Delta = \int_{-\infty}^{+\infty} \frac{\partial F_1} {\partial\theta} \phi(\theta) d \theta.
$$
\newpage
\section*{Appendix E: Results for Item Difficulty Recovery in Simulation Study 1}

The item difficulty estimates are reported in Table~\ref{s11numeric_delta}.

\begin{table}[!h]
\centering
\footnotesize
\caption{Bias and root mean square error (RMSE) for estimates of item difficulty in simulation study 1} 
\begin{tabular}{cccccccccc}
\hline
     &\multicolumn{2}{c}{TFM}& \multicolumn{2}{c}{GFM}& &\multicolumn{2}{c}{TFM}& \multicolumn{2}{c}{GFM}\\     
\cmidrule(r){2-3} \cmidrule(r){4-5}  \cmidrule(r){7-8} \cmidrule(r){9-10}
Item&	Bias&	RMSE&	Bias&	RMSE&	Item&	Bias&	RMSE&	Bias&	RMSE\\
\hline
01&	0.00&	0.08&	0.00&	0.08&	21&	-0.01&	0.06&	-0.01&	0.06\\
02&	0.00&	0.08&	0.00&	0.08&	22&	0.01&	0.07&	0.01&	0.07\\
03&	0.00&	0.08&	0.00&	0.08&	23&	0.00&	0.06&	0.00&	0.06\\
04&	0.00&	0.07&	0.00&	0.07&	24&	0.01&	0.07&	0.01&	0.07\\
05&	-0.01&	0.08&	-0.01&	0.08&	25&	0.01&	0.07&	0.01&	0.07\\
06&	0.01&	0.07&	0.00&	0.07&	26&	0.00&	0.07&	0.00&	0.07\\
07&	0.00&	0.07&	0.00&	0.07&	27&	0.00&	0.06&	0.00&	0.06\\
08&	0.00&	0.07&	0.00&	0.07&	28&	0.00&	0.06&	0.00&	0.06\\
09&	-0.01&	0.07&	-0.02&	0.07&	29&	0.00&	0.07&	0.00&	0.07\\
10&	0.00&	0.08&	0.00&	0.08&	30&	0.00&	0.07&	0.00&	0.07\\
11&	0.00&	0.07&	0.00&	0.07&	31&	0.00&	0.07&	0.00&	0.07\\
12&	0.00&	0.07&	0.00&	0.07&	32&	0.01&	0.06&	0.01&	0.06\\
13&	0.00&	0.07&	0.00&	0.07&	33&	0.00&	0.07&	0.01&	0.07\\
14&	0.00&	0.07&	0.00&	0.07&	34&	0.00&	0.07&	0.00&	0.07\\
15&	0.00&	0.08&	0.00&	0.08&	35&	0.00&	0.07&	0.00&	0.07\\
16&	0.00&	0.07&	0.00&	0.07&	36&	0.00&	0.06&	0.01&	0.06\\
17&	0.00&	0.07&	0.00&	0.07&	37&	0.00&	0.07&	0.01&	0.07\\
18&	-0.01&	0.07&	-0.01&	0.07&	38&	-0.01&	0.07&	0.00&	0.07\\
19&	0.00&	0.06&	0.00&	0.06&	39&	0.00&	0.07&	0.00&	0.07\\
20&	0.00&	0.06&	0.00&	0.06&	40&	-0.01&	0.07&	0.00&	0.07\\
\hline
\end{tabular}
\label{s11numeric_delta}
\end{table}
\newpage

\end{document}